\documentclass[doublecol]{epl2}
\usepackage{amsmath,bm,graphicx}
\usepackage{amssymb}
\usepackage{amsfonts}
\usepackage{color}
\usepackage[normalem]{ulem}
\newcommand{\be}{\begin{equation}}
\newcommand{\ee}{\end{equation}}
\newcommand{\bea}{\begin{eqnarray}}
\newcommand{\eea}{\end{eqnarray}}

\newcommand{\dd}[2]{\frac{d #1}{d #2}}

\title{Universality properties of steady driven coagulation with collisional evaporation}
\shorttitle{Universality properties of steady driven coagulation with collisional evaporation} %Insert here a short version of the title if it exceeds 70 characters

\author{Colm Connaughton\inst{1,2,3} \and Arghya Dutta\inst{4}\thanks{Present address: Universit\'{e} de Strasbourg, CNRS, Institut Charles Sadron, UPR 22, 67000 Strasbourg, France. e-mail: argphy@gmail.com} \and R. Rajesh\inst{4,5} \and Oleg Zaboronski\inst{2}}
\shortauthor{C. Connaughton \etal}
\institute{
\inst{1} Centre for Complexity Science, University of Warwick, Gibbet Hill Road, Coventry CV4 7AL, UK,\\
\inst{2} Mathematics Institute, University of Warwick, Gibbet Hill Road, Coventry CV4 7AL, UK,\\
\inst{3} London Mathematical Laboratory,  14 Buckingham St., London WC2N 6DF, UK,\\
\inst{4} Institute of Mathematical Sciences, CIT Campus, Taramani, Chennai-600113,India,\\
\inst{5} Homi Bhabha National Institute, Training School Complex, Anushakti Nagar, Mumbai 400094, India.}

\abstract{
Irreversible aggregation is an archetypal example of a system driven far from 
equilibrium by sources and sinks of a conserved quantity (mass). The source is 
a steady input of monomers and the evaporation of colliding particles with a 
small probability is the sink. Using exact and heuristic analyses, we find
a universal regime and two distinct non-universal regimes distinguished
by the relative importance of mergers between small and large particles. At the boundary between the regimes we find an analogue of the logarithmic correction 
conjectured by Kraichnan for two-dimensional turbulence.
}

\begin{document}

\maketitle

\emph{Introduction.} Understanding the diverse range of non-equilibrium statistical dynamics observed in physical systems with many interacting degrees of freedom relies heavily on identifying phenomena which occur frequently enough to make a unified theoretical treatment worthwhile. One example where such commonalities can be found is the case of systems in which microscopic interactions between degrees of freedom are constrained by conservation laws. Such systems can be driven into far-from-equilibrium steady states by the presence of sources and sinks of a conserved quantity. When these sources and sinks are widely separated, the physics of these systems is often controlled by steady conserved currents flowing between sources and sinks. Examples include fluid turbulence \cite{falkovich_particles_2001,boffetta_two-dimensional_2012}, wave kinetics \cite{newell_wave_2011}, granular gases \cite{ben-naim_stationary_2005} and irreversible aggregation \cite{leyvraz_scaling_2003,handbook}. In this work, as an archetypal example of the entire class, we study irreversible aggregation as its simplicity allows analytical treatment of the cascade
of the conserved quantities from source to sink~\cite{crz1,crz2}. We address the following fundamental questions: under what circumstances is the steady state of such a driven-dissipative system universal and what happens when it is not? Universality in this context means that the steady state becomes independent of the details of the source and sink when the separation between them tends to infinity. Universality is often assumed to hold for fluid turbulence. There are examples from wave kinetics however, such as Rossby wave turbulence, where it is known to fail \cite{connaughton_rossby_2015}.  However, a systematic analytical study of these questions is lacking.

In this work, we present an example from the kinetics of irreversible aggregation for which universality can be studied cleanly as the underlying microscopic dynamics encoded in the aggregation kernel, $K(i,j)$, is varied. The source is provided by steady input of small particles. The sink is provided by a mechanism which we call collisional evaporation, whereby  particles have a small probability, $\lambda$, of annihilating upon contact rather than merging. We study this mechanism primarily for convenience: it provides a sink for large particles which is analytically tractable. We were motivated, however, by recent work on the kinetics of fragmentation in planetary rings \cite{brilliantov_model_2009,brilliantov_size_2015} where a similar mechanism arises from physical considerations. In that context, large particles do not evaporate but fragment into small particles which remain in the system acting as an effective source. In addition to a known universal regime~\cite{hayakawa_irreversible_1987}, we 
find that there are two distinct non-universal regimes. Physically, these regimes are distinguished by the relative importance of mergers between very small and very large particles. We refer to this property as {\em locality} of interaction \cite{connaughton_stationary_2004} using terminology borrowed from the turbulence literature. At the boundary between regimes we find the analogue of the logarithmic correction conjectured by Kraichnan\cite{kraichnan_inertial-range_1971} in 1971 in the context of two-dimensional turbulence \cite{boffetta_two-dimensional_2012}.

\emph{Model and Results.} Let $N(m,t)$ be the density of particles of mass $m$. 
At the mean-field level, $N(m,t)$ evolves by a variant of the Smoluchowski kinetic equation:
\begin{align}
\label{ctssmol}
&\partial_t N(m,t)  =  - (1+\lambda)\int_0^\infty \!\!\! dm_1 N(m_1,t) N(m,t) K(m_1,m) \nonumber\\
&+\frac{1}{2} \int_0^m \!\!\!dm_1 N(m_1,t) N(m\!-\!m_1,t) K(m_1, m-m_1)\nonumber\\
&+\frac{J}{m_0} \delta (m-m_0), 
\end{align}
where $J$ is the mass input rate, $m_0$ is the mass of the monomer, $\lambda$ is the dimensionless collisional evaporation rate, and $K$ is the collision kernel. 
We consider the widely studied~\cite{leyvraz_scaling_2003} family of model kernels
\begin{equation}
K(m_1,m_2) = g\,(m_1^\mu m_2^\nu + m_1^\nu m_2^\mu), \label{eq:kernel}
\end{equation}
adopting the convention $\nu \geq \mu$ and including a constant, $g$, to provide dimensional consistency. In what follows it will be convenient to introduce the notation $\beta=\nu+\mu$ and $\theta=\nu-\mu$. Of physical interest is the limit of small $\lambda$. The dimensional parameters of the problem are $J$, $g$ and $m_0$. Taking limits $\lambda \to 0$ and $t \to \infty$ in Eq.~(\ref{ctssmol}), the universality hypothesis conjectures the existence of a steady state which is independent of $m_0$ for masses, $m \gg m_0$. If universality holds, then dimensional analysis implies the steady state scaling law 
\begin{equation}
\label{KolmogorovScaling}
N(m) \sim J^\frac{1}{2} g^{-\frac{1}{2}}\,m^{-\frac{\beta+3}{2}},
\end{equation}
where $N(m)$ is the steady state mass distribution.
The applicability of the universality assumption was established by Hayakawa \cite{hayakawa_irreversible_1987} who derived the exact asymptotic formula valid for $m \gg m_0$:
\begin{equation}
N(m) = \sqrt{\frac{J\,\left(1-\theta^2\right)\,\cos\left(\theta\,\pi/2\right)}{4\,\pi\,g}}\,m^{-\frac{\beta+3}{2}}, 
\label{eq:hayakawa}
\end{equation}
The derivation of this result is valid only for $\theta < 1$ giving a criterion for the applicability of the universality assumption. An analogous criterion holds for any scale-invariant kernel \cite{connaughton_stationary_2004}. 
To understand what happens when $\theta \geq 1$, the limits in Eq.~(\ref{ctssmol}) must be taken in the opposite order: $t \to \infty$ and then $\lambda \to 0$. The findings of such an analysis constitute the main results of this paper.
\begin{table}
\caption{\label{table1} Summary of results. 
The parameters $y$, $\tau_s$, $\eta_s$, $\tau_\ell$,
and $\eta_\ell$ are as defined in Eqs.~\eqref{ydefn}, \eqref{eq:smallmass},
and \eqref{eq:exponential}. The solution for $N(m)$  for integer $\theta$ may be found
in Eqs.~\eqref{result-integer} and \eqref{eq:borderlinesoln}.
}
%\begin{ruledtabular}
\begin{tabular}{l|l|l|l|l|l}
$\theta$&$y$& $\tau_s$ & $\eta_s$& $\tau_\ell$ & $\eta_\ell$\\
\hline
$0$ & $2$& $\frac{3 +\beta}{2}$ & $0$&$\frac{3 +\beta}{2}$&$0$\\
$(0,1)$ & $\frac{2}{\theta+1}$&  $\frac{3 +\beta}{2}$ & $0$&  $\frac{2 +\beta}{2}$ & $\frac{1}{2}$ \\
$(1,2)$ & $1$&$\frac{\beta+4-\theta}{2}$& $\frac{\theta-1}{2}$& $\frac{2 +\beta}{2}$ & $\frac{1}{2}$ \\
$(2,\infty)$ & $1$& $ \frac{\beta+\theta}{2}=\nu$ & $\frac{1}{2}$ & $ \frac{\beta+\theta}{2}=\nu$&$\frac{1}{2}$ 
\end{tabular}
%\end{ruledtabular}
\end{table}

The rescalings  $N(m) \to J^{1/2} g^{-1/2}\,m_0^{-(\beta+3)/2}\, N(m)$, $t \to J^{-1/2} g^{-1/2}\,m_0^{-(\beta-1)/2}\, t$ and $m \to m_0\,m$ remove all explicit dimensional parameters from Eq.~(\ref{ctssmol}). Exploiting the monodispersity of the source, it is convenient to work with the discrete form. 
In the steady state, setting the time derivative in Eq.~(\ref{ctssmol}) to zero, we obtain
\begin{align}
\label{smol}
0& =  \frac{1}{2} \sum_{m_1=1}^{m-1} 
\!\! N(m_1) N(m\!-\!m_1) K(m_1, m-m_1) \nonumber\\
&- (1+\lambda)\sum_{m_1=1}^\infty N(m_1) N(m) K(m_1,m) + \delta_{m,1}, 
\end{align}  
where $K(m_1,m_2) = m_1^\mu m_2^\nu + m_1^\nu m_2^\mu$  and $N(m)$ is the time independent
steady state mass distribution.
We assume that $N(m)$ has the following scaling form:
\be
N(m) = \frac{1}{m^{\tau}} f\left(\frac{m}{M} \right),~m, M \gg 1,
\label{eq:scaling1}
\ee
where $\tau$ is an exponent and $M$ plays the role of cutoff mass which diverges with $\lambda \rightarrow
0$ as
\be
M \sim \lambda^{-y}, \quad \lambda \to 0.
\label{ydefn}
\ee
In general, we expect different asymptotic behaviour for $m\ll M$ and $m \gg M$.
We define new exponents $\tau_s$, $\eta_s$, $\tau_{\ell}$ and $\eta_{\ell}$ which capture this:
\bea
N(m) &\simeq& \frac{a_s}{m^{\tau_s} M^{\eta_s}},~ m \ll M,
\label{eq:smallmass}\\
N(m) &\simeq& \frac{a_{\ell} e^{-m/M}}{m^{\tau_\ell} M^{\eta_\ell}} ,~ m \gg M,
\label{eq:exponential}
\eea
where the exponential decay with $m$  for large $m$ will be argued for through exact  solutions. 
Before delving into the technical details, the results obtained for the different exponents are summarized in Table.~\ref{table1}.

Continuity of $N(m)$ near $M$ leads to the exponent equality
\be
\tau_s+\eta_s=\tau_\ell+\eta_\ell.
\label{eq:equality1}
\ee
In general $\tau_\ell \neq \tau_s$ as is also seen in turbulence~\cite{sirovich_energy_1994}, where it is referred to as the ``bottleneck effect''. We use two related approaches to get information about the asymptotic behaviour: moment methods and generating function methods.  Given $\alpha \in \mathbb{R}$, the moment, $\mathcal{M}_\alpha$ and the associated generating function, $F_\alpha(x) $, are defined:
\begin{equation}
\mathcal{M}_\alpha = \sum_{m=1}^\infty m^\alpha\, N(m), ~~ F_\alpha(x) = \sum_{m=1}^\infty m^\alpha\, N(m)\, x^m.
\end{equation}
The two are  related by $\mathcal{M}_\alpha = F_\alpha(1)$.  

Generating function methods are based on analysing equations for $F_\alpha(x)$. They generally give information about the {\em large} mass asymptotics of $N(m)$.
Multiplying Eq.~(\ref{smol}) by $x^m$ and summing over all $m$, we obtain a relationship between $F_\nu(x)$ and $F_\mu(x)$: 
\be
\label{eq:f-new}
F_\nu(x)=\frac{(1+\lambda)\mathcal{M}_\nu F_\mu (x)-x}{F_\mu(x)-(1+\lambda)\mathcal{M}_\mu}.
\ee
A single equation for two unknown functions  does not allow us to determine $F_\nu(x)$ and $F_\mu(x)$. However, if $N(m)$ has the assumed form Eq.~(\ref{eq:exponential}) for large $m$, then $F_\mu(x)$ and $F_\nu(x)$ must have a singularity at a point $x_c$ on the positive real axis \cite{krapivsky_kinetic_2010}. The structure of this singularity is constrained by Eq.~(\ref{eq:f-new}). Generating function methods work by performing a consistency analysis of Eq.~(\ref{eq:f-new}) in the neighbourhood of $x_c$. This will allow us to  determine the exponents  $\tau_\ell$, $\eta_\ell$ characterizing the large mass behaviour of $N(m)$. Note that if $\nu = \mu+n$, where $n$ is a non-negative integer, then the relation $F_{\alpha+n} = \left(x d/dx \right)^n F_\alpha(x)$  can be used to close Eq.~(\ref{eq:f-new}):
\be
\label{eq1}
\left(x\dd{ }{x}\right)^n F_\mu(x)=\frac{(1+\lambda)\mathcal{M}_\nu F_\mu (x)-x}{F_\mu(x)-(1+\lambda)\mathcal{M}_\mu}.
\ee
This equation will provide some exact results for particular cases which do not rely on the assumptions underpinning the general singularity analysis.

In contrast to generating function methods, moment methods are based on analysing equations for $\mathcal{M}_\alpha$.   They generally give information about the {\em small} mass asymptotics of $N(m)$. Multiplying Eq.~(\ref{smol}) by $m^n$ and summing over $m$, we obtain a hierarchy of equations relating moments of different orders:
\begin{align}
&\lambda \left({\mathcal M}_{\mu} {\mathcal M}_{\nu+n}+
{\mathcal M}_{\mu+n} {\mathcal M}_{\nu} \right) =  \frac{-1}{2 \lambda+1} \delta_{n,0}+\nonumber \\
&\sum_{k=1}^{n-1}
{n \choose k} {\mathcal M}_{\mu+k} {\mathcal M}_{\nu+n-k} +1, \quad
n=0,1,2,\ldots
\label{eq:momentsgen}
\end{align}
In the limit $\lambda \to 0$, any particular moment is either dominated by the small mass cutoff, $m=1$, or by the large mass cutoff $M(\lambda)$ (setting aside marginal cases). Moment methods work by requiring these dependences to be consistent across the above hierarchy. Such consistency conditions put constraints on the small mass behaviour of $N(m)$ which will allow us to obtain information about the exponents $\tau_s$, $\eta_s$.

\emph{Exact analysis.}
When $\mu=\nu$, the model may be solved exactly. Eq.~(\ref{eq:f-new}) 
reduces to a quadratic equation in $F_\mu(x)$ that is satisfied by
\be
F_\mu(x) = (1+\lambda) {\mathcal M}_\mu - \sqrt{(1+\lambda)^2 {\mathcal M}_\mu^2 - Jx}.
\label{eq:nueqmuexact}
\ee
Determining the coefficient of $x^m$, we obtain
\be
N(m) \simeq
\sqrt{\frac{1}{4 \pi}}
\frac{1}{m^{\frac{3 + \beta}{2}}} e^{-m /M}, ~m, M \gg 1,
\label{eq:mu=nu}
\ee
where
$M= \lambda^{-2},  \lambda\rightarrow 0$. This solution is valid for both $m\ll M$ and $m \gg M$.
In the limit 
$\lambda\to 0$, the result Eq.~\eqref{eq:mu=nu} coincides with the result for the sink at 
infinity [see Eq.~\eqref{eq:hayakawa}].

We also obtain exact results when $\theta=\nu-\mu$ is an integer, in which case the generating
functions satisfy Eq.~\eqref{eq1}.
At the singularities  in the complex $x$-plane,
the coefficient of the highest order term is zero.
Therefore, at  the singular point $x_c$, $F_\mu$ satisfies 
\begin{eqnarray}
F_{\mu}(x_c)=(1+\lambda){\mathcal M}_\mu.
\label{eq17}
\end{eqnarray}
This relation allows us to expanding $F_{\mu}(x)$ about $x_c$. 
Doing a careful analysis of the
singular terms, the details of which will be published elsewhere, we obtain
\be
N(m) \simeq
\begin{cases}
\frac{m^{-(2+\beta)/2}}{\sqrt{2\pi M}} e^{-m/M}, & \theta=1, \\
\frac{m^{-\nu}}{\sqrt{2M}}
\frac{e^{-m/M}}{\sqrt{\ln m}}, & \theta=2,\\
\frac{m^{-\nu}}{M F_{\mu+1}(x_c)} e^{-m/M}, &\theta=3,4,\ldots,
\end{cases}
\label{result-integer}
\ee
for $m\gg M$.
We observe that logarithmic corrections to the power law appear only for $\theta=2$. 
Also, for $\theta\geq 2$, $\tau_\ell$, the exponent
characterizing the power law remains equal to $\nu$,
independently of $\theta$.

This leads us to consider the exact solution of
a simplified model that reproduces the correct exponents for $\theta>2$.
For such $\theta$, mass transfer from small to large masses is expected to be dominated by 
collisions between large and small masses. 
This aspect is captured by the so-called addition model \cite{Hendriks1984,brilliantov_nonscaling_1991} where only
coagulations that  involve at least one particle of mass one are allowed such that
the collision kernel is
$K(m_1,m_2) = (m_1^\mu m_2^\nu + m_1^\nu m_2^\mu) (\delta_{m_1,1}+
\delta_{m_2,1})$. In this kernel, the terms in $\mu$ are subdominant, and the kernel
may be rewritten as $K(m_1,m_2) = m_1^\nu \delta_{m_2,1}+ m_2^\nu \delta_{m_1,1}$,
such that the resultant $N(m)$ should not depend on $\mu$.
By substituting into the Smoluchowski
equation [Eq.~\eqref{smol}], it is straightforward to solve for the mass distribution $N(m)$.
In the limit $\lambda \to 0$, we obtain
\be
N(m) \approx \frac{\sqrt{2 J} e^{-m/M}}{m^\nu \sqrt{M}}, ~m \gg
1,~M \to \infty,
\ee
where 
$M= \lambda^{-1} [1+O(\lambda)]$. 
Note that the exponent $\tau_\ell=\nu$ coincides with that obtained in Eq.~\eqref{result-integer} 
for $\theta=2,3,\ldots$. We thus expect that $\eta_\ell=1/2$ for these values of $\theta$.

\emph{Analysis of singularities.} 
The large mass behaviour for non-integer values of $\theta$ may be determined by
analysing  Eq~\eqref{eq:f-new} near
the singular point. Let the singularity of $F(x)$ closest to the origin be denoted by
$x_c=e^{1/M}$. 
Consider $x=x_c-\epsilon$, $\epsilon \to 0^+$. For $N(m)$  as in
Eq.~\eqref{eq:exponential},  the leading singular behaviour of the
generating functions  $F_\nu$ and $F_\mu$  is proportional to
$\epsilon^{\tau-\nu-1}$ and
$\epsilon^{\tau-\mu-1}$ respectively. 
We now claim that $F_\mu(x_c) = (1+\lambda)\mathcal{M}_\mu$. Suppose this was not the case and
$F_\mu(x_c) \neq (1+\lambda)\mathcal{M}_\mu$. Then, by expanding about $x_c$, it follows from 
Eq.~\eqref{eq:f-new} that  $F_\mu(x)$ and $F_\nu(x)$ would have same singularity near $x=x_c$. 
This implies that $\mu=\nu$. For this case, from the exact solution,
it is easily seen that   that $F_\mu(x_c) = (1+\lambda)\mathcal{M}_\mu$, leading
to a contradiction.
When $\mu\neq\nu$,
$F_\mu(x)$ and $F_\nu(x)$ should have different singular behaviour near $x=x_c$, again leading
to a contradiction. We therefore
conclude that $F_\mu(x_c)=(1+\lambda)\mathcal{M}_\mu$, as also seen in Eq.~\eqref{eq17} for integer  $\theta$.  
This in conjunction with $n=1$ in Eq.~\eqref{eq:momentsgen} implies that the leading term in the
numerator of Eq.~\eqref{eq:f-new} is $-M^{-1}$.
We now expand the generating functions about $x=x_c$ as
\begin{align}
&F_\mu(x_c\!-\!\epsilon)=   (1+\lambda)\mathcal{M}_\mu- \epsilon^{\tau_\ell-\mu-1} R_1(\epsilon) -
\epsilon R_2(\epsilon), \label{eq:fsingularIII}\\
&F_\nu(x_c\!-\!\epsilon) =  \epsilon^{\tau_\ell-\nu-1} R_3(\epsilon) + \epsilon R_4(\epsilon),\label{eq:fsingularIV}
\end{align}
where $R_i$'s are regular in $\epsilon$, $R_1(0) \neq 0$,
$R_3(0) \neq 0$ and $\tau_\ell<\nu+1$. 
Substituting into Eq.~\eqref{eq:f-new},
we obtain
\be
\label{eq:sing-main}
\epsilon^{\tau_\ell-\nu-1} R_3(\epsilon) + \epsilon R_4(\epsilon)=\frac{M^{-1}+O(\epsilon)}{\epsilon^{\tau_\ell-\mu-1} R_1(\epsilon) +
\epsilon R_2(\epsilon)}.
\ee
We now compare the leading singular behaviour on the both sides of Eq.~\eqref{eq:sing-main}.

First, when $0<\tau_\ell-\mu-1 <1$,  the denominator of Eq.~\eqref{eq:sing-main}  is dominated by 
$\epsilon^{\tau_\ell-\mu-1} R_1(\epsilon)$, and by comparing the leading
singular terms on both sides of Eq.~\eqref{eq:sing-main},
we obtain 
\be
\tau_\ell=\frac{\beta+2}{2}, \quad 0<\theta<2\label{eq:tau-l-local},
\ee
where  the constraint on $\theta$ follows from our assumption $0<\tau_\ell-\mu-1 <1$.
Comparing  the coefficients of the leading singular terms we obtain
$R_3(0) R_1(0)= M^{-1}$, $M \to \infty$.
Knowing $R_1(0)$, $R_3(0)$, we perform an inverse Laplace transform to obtain
\be
N(m) \simeq \sqrt{\frac{\theta \sin\frac{\pi\theta}{2}}{2 \pi M}} 
\frac{e^{-m/M}}{m^{(2+\beta)/2}},~ m \gg M, ~0<\theta<2.
\ee

Second, consider the  case $\tau_\ell-\mu-1 >1$. The denominator of Eq.~\eqref{eq:sing-main} 
is dominated by $\epsilon R_2(\epsilon)$.   Comparing the leading singular terms on both sides of  
Eq.~\eqref{eq:sing-main}, we obtain
\be
\tau_\ell=\frac{\beta+\theta}{2}=\nu, \quad \theta>2, 
\label{eq:tau-l-addition}
\ee 
where  the constraint on $\theta$ follows from our assumption $\tau_\ell-\mu-1 >1$ .
Comparing the coefficients of the leading singular terms we obtain
$R_2(0) R_3(0)= M^{-1}$, $M \to \infty$.
Doing an inverse Laplace transform, we obtain
\be
N(m) \simeq \frac{m^{-\nu}}{MF_{\mu+1}(x_c)} e^{-m/M}, ~m \gg M,~\theta>2,
\ee
where we used $R_2(0)=F_{\mu+1}(x_c)$. It is also straightforward to show that
$F_{\mu+1} (x_c) \sim M^{-\min (\eta_\ell+\theta-2, \eta_s)}$  for $\theta>2$, allowing us to obtain
$\eta_s+\eta_\ell=1$. However, from the results for the addition model, we know
that $\eta_\ell =1/2$ for $\theta>2$. Thus,
\be
\eta_s=\frac{1}{2}; ~~\eta_\ell=\frac{1}{2},~\theta>2.
\label{eq:equality}
\ee

\emph{Moment analysis.}
The exponents describing the small mass behaviour of the mass
distribution $N(m)$ [see Eq.~\eqref{eq:smallmass}] may be determined
using the relations between the first three moments of the mass [see Eq.~\eqref{eq:momentsgen}].
By determining when the integrals diverge at large masses, we obtain 
${\mathcal M}_\alpha \sim  M^{-\eta_s} M^{\max(\alpha+1-\tau_s,0)}$ when $\alpha \neq
\tau_s-1$ and ${\mathcal M}_\alpha \sim M^{-\eta_s} \ln M$ when $\alpha=\tau_s-1$.
It is straightforward to obtain some simple bounds for the exponents. { We start by writing down 
explicitly the  equations for $n=0,1,2$ in Eq.~\eqref{eq:momentsgen}:
\begin{subequations}
\label{eq:moments}
\bea
{\mathcal M}_\mu {\mathcal M}_\nu &=& \frac{J}{2 \lambda +1},  \label{eq:nstar} \\
\lambda \left({\mathcal M}_{\mu} {\mathcal M}_{\nu+1}\!+\! {\mathcal M}_{\mu+1}
{\mathcal M}_{\nu} \right)\!\!& = & J,
\label{eq:m1star} \\
\lambda \left({\mathcal M}_{\mu} {\mathcal M}_{\nu+2}\!+\! {\mathcal M}_{\mu+2}
{\mathcal M}_{\nu} \right)\!\!& = &  2 {\mathcal M}_{\mu+1}
{\mathcal M}_{\nu+1}\!\!+ \!J. \label{eq:m2star}
\eea
\end{subequations}
The moments ${\mathcal M}_x$ depend on the upper cutoff $M$ as
\be
{\mathcal M}_x \sim \int^M dm \frac{a_s M^{-\eta_s}}{m^{\tau_s-x} },
\label{eq:integral}
\ee
where $x \sim y$ means that $x/y$ is
$O(M^0)$ when $\lambda \to 0$.
Clearly,
\begin{subequations}\label{eq:momentdiv}
\bea
{\mathcal M}_x &\sim&  M^{-\eta_s} M^{\max(x+1-\tau_s,0)},~x\neq
\tau_s-1,
\label{eq:momentdiv1}\\
&\sim& M^{-\eta_s} \ln M,~x= \tau_s-1,
\label{eq:momentdiv2}
\eea
\end{subequations}
for any $x$. 

We first show that $\eta_s \geq 0$. Suppose $\eta_s<0$. Then all moments of
$m$ diverge as $M\to \infty$ [see Eq.~(\ref{eq:momentdiv})]. 
Since ${\mathcal M}_\nu$ and ${\mathcal M}_\mu$ both diverge, 
Eq.~(\ref{eq:nstar}) has no solution. Therefore,  $\eta_s \geq 0$.

Next, we derive upper and lower bounds for the exponent $\tau_s$. We
first show that $\tau_s < \nu+2$. Suppose $\tau_s > \nu+2$. Then,
from Eq.~(\ref{eq:momentdiv}), 
${\mathcal M}_\mu \sim {\mathcal M}_{\mu+1}\sim {\mathcal M}_{\nu}\sim {\mathcal M}_{\nu+1} 
\sim M^{-\eta_s}$.  From Eq.~(\ref{eq:nstar}), we immediately obtain
$\eta_s=0$.  The left hand side of  Eq.~(\ref{eq:m1star}) is dominated by
the first term such that $\lambda{\mathcal M}_{\nu+1}\sim J$. 
This implies that $\lambda \sim O(1)$. But $\lambda$ is a parameter
that tends to zero. Hence, there is a contradiction and we conclude that 
$\tau_s \leq \nu+2$. 

Now consider $\tau_s = \nu+2$. Now from Eq.~(\ref{eq:momentdiv})
${\mathcal M}_\mu \sim {\mathcal M}_{\mu+1}\sim {\mathcal M}_{\nu}\sim M^{-\eta_s}$, ${\mathcal M}_{\nu+1}\sim M^{-\eta_s}\ln M$ and ${\mathcal M}_{\nu+2} 
\sim M^{1-\eta_s}$. From Eq.~\eqref{eq:nstar}, we immediately obtain $\eta_s=0$. Since the left hand side of Eq.~\eqref{eq:m1star} is dominated by the first term we obtain $\lambda\sim(\ln M)^{-1}$. The left hand side of Eq.~\eqref{eq:m2star} is dominated by the first term such that $\lambda{\mathcal M}_{\nu+2}\sim {\mathcal M}_{\nu+1}$ or $\lambda\sim(\ln M/M)$, leading to a contradiction. Hence $\tau_s\neq \nu+2$, and we conclude that $\tau_s<\nu+2$.

Second, we show that $\tau_s \geq
\mu+1$. Suppose $\tau_s<\mu+1$. From
Eq.~(\ref{eq:momentdiv}), we obtain that the integrals for ${\mathcal
M}_{\mu}$ and ${\mathcal M}_\nu$ diverge with $M$ as
${\mathcal M}_{\mu} \sim M^{\mu+1-\tau_s-\eta_s}$ and
${\mathcal M}_{\nu} \sim M^{\nu+1-\tau_s-\eta_s}$. Also
${\mathcal M}_{\mu+1} \sim M {\mathcal M}_{\mu}$ and
${\mathcal M}_{\nu+1} \sim M {\mathcal M}_{\nu}$. 
Eq.~(\ref{eq:m2star}) reduces to 
$\lambda  M^2 {\mathcal M}_{\mu}{\mathcal M}_{\nu} \sim M^2  
{\mathcal M}_{\mu} {\mathcal M}_{\nu}$ or $\lambda \sim O(1)$. Since $\lambda$ is a parameter that tends to zero, we obtain a contradiction here. Hence, $\tau_s  \geq \mu+1$. Combining the bounds, we obtain
\be
\label{eq:tau_s_bound}
\mu+1\leq \tau_s < \nu +2; ~\quad \eta_s \geq 0.
\ee
}

Given these bounds, Eq.~\eqref{eq:moments} may be rewritten as
\be
{\mathcal M}_\mu {\mathcal M}_\nu \sim  1; ~
\frac{{\mathcal M}_{\nu}}{ {\mathcal M}_{\nu+1}}
\sim \lambda;~
{\mathcal M}_{\mu+1}
{\mathcal M}_{\nu+1} \sim  M.
\label{eq:momentsnew}
\ee
Substituting for ${\mathcal M}_\alpha$ in terms of $\tau_s$, $\eta_s$, and $y$ and comparing
the exponents, we obtain
\begin{subequations}
\label{eq:moments-new}
\bea
2\eta_s &=& \max(\nu+1-\tau_s,0)\label{eq:momentsnew-1},\\
\frac{1}{y} &=& (\nu+2-\tau_s)-\max(\nu+1-\tau_s,0),\label{eq:momentsnew-2}\\
2\eta_s &=& \nu+1-\tau_s+\max(\mu+2-\tau_s,0)\label{eq:momentsnew-3}.
\eea
\end{subequations}

By considering whether $\tau_s$ is greater or less than $\nu+1$, it is straightforward to find the solution
to Eq.~\eqref{eq:moments-new} to be 
$\tau_s = (3+\beta)/2$,
$\eta_s = 0$, 
$y =2/(\theta+1)$  for local kernels ($\theta<1$), and
$\eta_s =(\nu+1-\tau_s)/2$, 
$y =1$ for non-local kernels ($\theta>1$). By combining these results with
those for the large mass asymptotics from the analysis of singularities through
the exponent equality Eq.~\eqref{eq:equality1}, we are able to solve for
all the exponents as summarized in Table~\ref{table1}.

We now examine the case when $\theta=\nu-\mu=1$, the boundary between
the local and non-local kernels when logarithmic corrections to the power law prefactors
are expected. We assume 
the following form for $N(m)$:
\be
N(m) \sim \frac{ (\ln m)^{-x} (\ln M)^{-z}}{m^{\nu+1}
M^{\eta_s}},~m \ll M,~\theta=1,
\label{eq:borderline}
\ee
where $x$ and $z$ are new exponents characterizing the logarithmic
corrections. In addition, the cutoff mass scale could depend logarithmically on 
$\lambda$. It is then straightforward to obtain 
${\mathcal M}_\mu \sim M^{-\eta_s} (\ln M)^{-z}$, 
${\mathcal M}_{\mu+1} \sim {\mathcal M}_\nu
\sim M^{-\eta_s} (\ln M)^{-z+\max(0,1-x)}$ if $x\neq 1$, and 
${\mathcal M}_{\nu+1} \sim M^{1-\eta_s} (\ln M)^{-x-z}$. 
By substituting these expressions into Eq.~\eqref{eq:momentsgen} for $n=0,1,2$
and looking for a consistent solution, we obtain $\eta_s=0$,
$x=0$ and $z=1/2$. Thus, we obtain that the mass distribution takes the form
\bea
N(m) &\sim& \frac{1 }{m^{1+\nu} \sqrt{\ln M}},~m \ll M, ~\theta=1,\\
M&\sim & \frac{\ln \lambda}{\lambda}, ~\lambda \to 0, ~\theta=1.
\label{eq:borderlinesoln}
\eea

\emph{Summary.} To summarize, we have combined exact solutions and scaling heuristics to fully characterize the steady state of 
irreversible coagulation with constant input of monomers and removal of large particles by collisional evaporation. 
The technical results are summarized in Table~\ref{table1}. Conceptually, the most important finding is that in contrast to a priori expectations, 
there are two distinct non-local regimes corresponding to $1 < \theta < 2$ and $\theta>2$. 
In the first regime, the mass distribution $N(m)$ retains a dependence on the sink scale $M$ but becomes independent of the source scale, $m_0$. 
In the latter regime, $N(m)$ depends on both source and sink. Logarithmic corrections are found at the boundaries between regimes. 
These are analogous to the correction proposed by Kraichnan~\cite{kraichnan_inertial-range_1971} to account for the marginal non-locality of the 
enstrophy cascade in two-dimensional turbulence. In a forthcoming publication, we will provide numerical evidence for the assumptions 
underpinning the scaling analysis and assess the extent to which the predicted logarithmic corrections can be measured.
The question remains as to how sensitive the two non-local regimes are to the nature of the sink. Answering this
would require a detailed numerical study of different sinks. There is some evidence for the sink independence of the regime
where $N(m)\sim m^{-\nu}$, as the same scaling was obtained for the model with a hard cutoff [evaporation of particles
larger than a cutoff mass]~\cite{ball_instantaneous_2011}. The other non-local regime $1<\theta<2$ is less explored.
Studying the robustness of these regimes is a promising area for future studies.
 
In this paper, we  studied the steady state but not the dynamics leading to it. There are good reasons to consider this in the future. 
It is often the case that the system does not reach a steady state, as assumed in this paper. 
For instance, for kernels of the form $m_1^\mu + m_1^\nu$ with
$\mu <-1$, and in the absence of collision-dependent evaporation, it is known that the system does not reach a 
stationary state at large times~\cite{mendes1999}. It would be interesting to see if evaporation induces a steady state.
Even in the local case, $\theta <1$, the dynamics leading to the steady state must be very different for gelling ($\beta>1$) and non-gelling ($\beta<1$)
kernels. Furthermore, in the non-local case, evidence from closely 
related models~\cite{ball_collective_2012,ball_instantaneous_2011} suggests that the steady state could become unstable for 
$\lambda\to 0$. Such an instability would result in persistent oscillatory kinetics. The relative tractability of the collisional evaporation model 
presented here may facilitate analytic study of this stability question.

\acknowledgments
CC, AD \& RR acknowledge funding from the LMS (No. 41238) and the EPSRC (No. EP/M003620/1). CC \& AD also acknowledge the hospitality of the International Centre for Theoretical Sciences (ICTS) during a visit for participating in the program ``Non-equilibrium statistical physics" (Code: ICTS/Prog-NESP/2015/10) where part of the manuscript was written.

%\bibliographystyle{eplbib}
%\bibliography{references}

\begin{thebibliography}{10}
\expandafter\ifx\csname url\endcsname\relax\def\url#1{\texttt{#1}}\fi

\bibitem{falkovich_particles_2001}
\Name{Falkovich G., Gawedzki K. \and Vergassola M.} \REVIEW{Rev. Mod. Phys.}{73}{2001}{913}.

\bibitem{boffetta_two-dimensional_2012}
\Name{Boffetta G. \and Ecke R.~E.} \REVIEW{Ann. Rev. Fluid Mech.}{44}{2012}{427}.

\bibitem{newell_wave_2011}
\Name{Newell A.~C. \and Rumpf B.} \REVIEW{Ann. Rev. Fluid Mech.}{43}{2011}{59}.

\bibitem{ben-naim_stationary_2005}
\Name{Ben-Naim E. \and Machta J.} \REVIEW{Phys. Rev. Lett.}{94}{2005}{138001}.

\bibitem{leyvraz_scaling_2003}
\Name{Leyvraz F.} \REVIEW{Phys. Rep.}{383}{2003}{95}.

\bibitem{handbook}
\Name{Connaughton C., Rajesh R. \and Zaboronski O.} \Book{Handbook of Nanophysics: Clusters and Fullerenes}
\Editor{K. D. Sattler}\Publ{Taylor and Francis, Boca Raton}\Year{2010}

\bibitem{crz1}
\Name{Connaughton C., Rajesh R. \and Zaboronski O.} \REVIEW{Phys. Rev. Lett.}{94}{2005}{194503}.

\bibitem{crz2}
\Name{Connaughton C., Rajesh R. \and Zaboronski O.} \REVIEW{Phys. Rev. Lett.}{98}{2007}{080601}.


\bibitem{connaughton_rossby_2015}
\Name{Connaughton C., Nazarenko S. \and Quinn B.} \REVIEW{Phys. Rep.}{604}{2015}{1}.

\bibitem{brilliantov_model_2009}
\Name{Brilliantov N.~V., Bodrova A.~S. \and Krapivsky P.~L.} \REVIEW{J. Stat. Mech.}{6}{2009}{11}.

\bibitem{brilliantov_size_2015}
\Name{Brilliantov N. \etal} \REVIEW{Proc. Nat. Acad. Sci.}{112}{2015}{9536}.

\bibitem{hayakawa_irreversible_1987}
\Name{Hayakawa H.} \REVIEW{J. Phys. A}{20}{1987}{L801}.

\bibitem{connaughton_stationary_2004}
\Name{Connaughton C., Rajesh R. \and Zaboronski O.} \REVIEW{Phys. Rev. E}{69}{2004}{061114}.

\bibitem{kraichnan_inertial-range_1971}
\Name{Kraichnan R.~H.} \REVIEW{J. Fluid. Mech.}{47}{1971}{525}.

\bibitem{sirovich_energy_1994}
\Name{Sirovich L., Smith L. \and Yakhot V.} \REVIEW{Phys. Rev. Lett.}{72}{1994}{344}.

\bibitem{krapivsky_kinetic_2010}
\Name{Krapivsky P., Redner S. \and Ben-Naim E.} \Book{A Kinetic View of Statistical Physics} (Cambridge University Press, Cambridge) 2010.

\bibitem{Hendriks1984}
\Name{Hendriks E. \and Ernst M.} \REVIEW{J. Coll. Inter. Sci.}{97}{1984}{176 }.

\bibitem{brilliantov_nonscaling_1991}
\Name{Brilliantov N.~V. \and Krapivsky P.~L.} \REVIEW{J. Phys. A}{24}{1991}{4789}.
  
\bibitem{mendes1999}
\Name{Krapivsky P.~L., Mendes J.~F.~F. \and Redner S.} \REVIEW{Phys. Rev. B}{59}{1999}{15950}.

\bibitem{ball_collective_2012}
\Name{Ball R.~C., Connaughton C., Jones P.~P., , Rajesh R. \and Zaboronski O.}\REVIEW{Phys. Rev. Lett.}{109}{2012}{168304}.

\bibitem{ball_instantaneous_2011}
\Name{Ball R.~C., Connaughton C., Stein T. H.~M. \and Zaboronski O.}\REVIEW{Phys. Rev. E }{84}{2011}{011111}.

\end{thebibliography}

\end{document}